\documentclass[10pt,a4paper]{article}
\usepackage{amsfonts,amssymb}
\usepackage[dvips]{graphicx}
\DeclareGraphicsExtensions{.eps}
\usepackage{epsfig}
\usepackage{graphicx}
\font\twelvei = cmmi10 scaled\magstep1
       \font\teni = cmmi10 
\font\mbf = cmmib10 scaled\magstep1
       \font\mbfs = cmmib10 \font\mbfss = cmmib10 scaled 833
\font\msybf = cmbsy10 scaled\magstep1
       \font\msybfs = cmbsy10 \font\msybfss = cmbsy10 scaled 833
\def\lsim{\raise0.3ex\hbox{$<$}\kern-0.75em{\lower0.65ex\hbox{$\sim$}}}
\def\gsim{\raise0.3ex\hbox{$>$}\kern-0.75em{\lower0.65ex\hbox{$\sim$}}}

\title
{Hydrodynamic Simulation of Two Component Advective Flows around Black Holes}
\author
{Kinsuk Giri\thanks{kinsuk@bose.res.in}$^{1}$, Sandip K. Chakrabarti\thanks{chakraba@bose.res.in}$^{1,2}$\\
$^{1}$S. N. Bose National Centre for Basic Sciences, Salt Lake, Kolkata 700098, India\\
$^{2}$Indian Centre for Space Physics, Chalantika 43, Garia Station Rd., Kolkata, 700084, India}
\begin{document}

\date{}


\maketitle

\label{firstpage}

\begin{abstract}
We carry out a series of numerical simulations of viscous accretion flows having a    
reasonable spatial distribution of the viscosity parameter. We add the power-law cooling throughout the flow. We show that, in agreement 
with the theoretical solutions of viscous transonic flows, matter having the viscosity parameter
above a critical value becomes a Keplerian disk while matter having lesser viscosity
remains a low angular momentum, sub-Keplerian flow. The latter component produces centrifugal pressure 
supported shock waves. Thus, for instance, a flow having sufficiently high viscosity 
on the equatorial plane and low viscosity above and below, would produce a Two Component Advective Flow
(TCAF) where a Keplerian disk is surrounded by a rapidly infalling sub-Keplerian halo. We find that the 
post-shock region of the relatively cooler Keplerian disk is evaporated and the overall configuration is quite stable. 
This agrees with the theoretical model with two components which attempt to explain the spectral and timing properties of 
black hole candidates.

\end{abstract}


\section{Introduction}
The problem of understanding the black hole accretion started with the discovery of quasars
back in the 1960s (Hazard et al. 1963; Schmidt, 1963). The explanation with a Bondi flow (Bondi 1952) solution was
quickly dismissed because of its low efficiency. The efficiency is higher in a
standard Keplerian disk (Shakura \& Sunyaev, 1973, hereafter SS73) which was also successful in explaining the 
so-called big blue bump of the quasars (Malkan et al. 1982).
However, observation of non-thermal photons in the spectrum (Sunyaev, \&  Truemper, 
1979) prompted the model builders to imagine that a hot electron cloud (the so-called 
Compton cloud) along with the standard disk could resolve the issue 
(Sunyaev \& Titarchuk, 1980, 1985). Numerous avatars
of the illusive Compton cloud  are present in the literature (e.g., Zdziarski, 1988 ; Haardt et al., 1994 ; Chakrabarti
\&  Titarchuk 1995, hereafter CT95). CT95, based on the solutions of viscous and inviscid transonic flows
around black holes (Chakrabarti, 1989, hereafter C89; Chakrabarti, 1990) proposed that, in general,
the accretion disk should really have two components: a Keplerian accretion on the equatorial plane and 
a sub-Keplerian inflow which surrounds the Keplerian disk, and the puffed up inner part of this latter component
is nothing but the Compton cloud. While the two component advective disk (TCAF) solution of CT95
was able to explain the spectral and timing properties including time lags observed in several black hole
candidates (Wu et al, 2002; Smith et al 2001a, 2001b, 2002, 2007; Rao et al, 2000), there is as yet no work in the literature to
show that the TCAF solution possible and if so, whether it is stable. The cause for concern was obvious: a Keplerian disk is 
necessarily sub-sonic, while the sub-Keplerian flow is supersonic, and becomes sub-sonic
only at the shock wave. The region between the shock-wave and the
sonic point near the horizon is known as the CENtrifugal pressure supported BOundary Layer or CENBOL, which in CT95 acts as the Compton cloud.
The questions remained unanswered are: (a) Under what circumstances TCAF actually forms? 
(b) Would the Keplerian component remain stable when the sub-Keplerian component flies by or it is disrupted or evaporated?
(c) What is the fate of the inner Keplerian disk component when the hot CENBOL actually forms? 
(d) How would the subsonic Keplerian disk and the sub-sonic CENBOL interact, and finally
(e) In presence of both the components what would be the net angular momentum distribution of the flow? 

In the present paper, we will address these vital issues. Through numerical simulations of 
viscous accretion flow with a power-law cooling effects, we show that only when the injected 
sub-Keplerian flow angular momentum is high enough and/or the viscosity is high enough, 
TCAF would be formed, otherwise the sub-Keplerian flow would remain sub-Keplerian. 
The TCAF, when formed, is a stable configuration. We would also show that the hot CENBOL 
effectively removes the Keplerian disk from the equatorial plane and the disk becomes truncated.

One of the earliest numerical simulations to study the behaviour of non-viscous flows around black holes 
was made nearly three decades ago (Hawley et al. 1984a, 1984b). The results clearly 
showed existence of shock waves slowly receding away from the black hole.     
Taam et al. (1991) considered a 2D axisymmetric accretion flow including momentum 
deposition by radiation as well as radiative heating and cooling. They found that the flow can be
unsteady. Matsuda et al. (1992) showed that 2D adiabatic flows are unstable to the 'flip-flop' instability for
accretion from a homogeneous medium and the density or velocity gradients merely 
provide an initial perturbation. Ishii et al. (1992) found that 2D isothermal flows exhibit the 'flip-flop' instability both
in the homogeneous and non homogeneous cases. Chakrabarti \& Molteni (1993) presented results of Smoothed Particle 
Hydrodynamics (SPH) simulations of thin accretion flows and winds. They showed that the
results of the simulation agree with the theoretical work (C89) on
the shock formation. The most significant conclusion was that the shocks in an inviscid flow were stable.
Molteni, Lanzafame \& Chakrabarti (1994, hereafter MLC94) simulated the formation of
thick disk by the SPH code and showed that the stable shocks do form here as well.
Molteni, Sponholz \& Chakrabarti (1996) included cooling effects and showed that when the 
cooling time scale roughly agrees with the infall time scale, the shocks oscillate.
Chakrabarti \& Molteni (1995, hereafter CM95) studied the numerical evolution 
of viscous isothermal disks, and showed that in one dimensional simulations, the 
shock wave goes away from the black hole. Igumenshchev et al. (2000) studied the dependence on the polytropic index 
$\gamma$ which varied from $4/3$ to $5/3$ as well as viscosity parameter 
$\alpha$ and found that the stability of the solutions depend on these parameters.
In Igumenshchev \& Abramowicz 1999, 2000; Stone, Pringle \& Begelman 1999; Proga \& Begelman 2003,
etc. the authors do not find shocks simply because they start and/or inject with Keplerian or quasi-Keplerian flows which 
are subsonic and are incapable for forming shocks. 
Those simulations were viscous, so a Keplerian flow remained Keplerian until very close to the black hole.
On the contrary the accretion flows  considered by Chakrabarti and collaborators are 
advective with a significant radial velocity component and the flow has angular momentum
much less than that of the Keplerian disk.

In the first paper of this series (Giri et al. 2010, hereafter Paper I) simulated the inviscid flow with low angular momentum
in two dimensions. In Giri \& Chakrabarti, 2012 (hereafter Paper II) 
the viscous flows were used to verify earlier predictions. Chakrabarti (1990) predicted that 
standing shocks are possible only when the viscosity parameter is less than a critical value ($\alpha_{crit}$ 
which depends on flow parameters). For $\alpha > \alpha_{crit}$,
the shock would move outward and the disk would become sub-sonic and Keplerian. 
In Paper II, we see precisely this. We show that when the viscosity parameter is very small
($\alpha < \alpha_{crit}$), the transport rates of
angular momentum on both sides of the shock could match and the shock can
remain steady. In this case, the shock becomes weaker as compared to an inviscid flow and forms farther away from
the black hole. However, when $\alpha > \alpha_{crit}$, the shock wave
is driven outwards and the disk between the shock and sonic point 
(located outside the horizon at about $r\sim 2.5$ Schwarzschild radii)
becomes Keplerian. At the same time, the viscous flow must also be cooled in order that not only the
flow has the Keplerian angular momentum distribution but also its temperature is cold enough 
so that a standard (Shakura-Sunyaev, 1973) disk is produced.  

In the present paper, we carry out numerical simulation in presence of both 
viscous and cooling effects. Earlier, Chakrabarti \& Das (2004) studied the effects of viscosity 
and showed that the shocks move outward with the increase in viscosity.
When cooling effects were also added, Das \& Chakrabarti (2004) showed that the 
shock can move inward under the combined effects of heating and cooling. Physically,
viscous heating increases post-shock pressure and also transports angular momentum faster.
As a result, the Rankine-Hugoniot condition is satisfied away from the black hole. Cooling, on the other hand,
reduces the post-shock pressure and the shock moves inward. With the combined effects, the shock
may or may not move outward. Since dwarf novae outbursts are known to be driven by enhanced viscosity on the 
equatorial plane (e.g., Cannizzo et al. 1982, 1995), we assume that the flows in general has this property. 
Thus, instead of using a constant $\alpha$ parameter throughout the simulation grid, 
we assume that $\alpha$ is maximum on the equatorial plane
and gradually goes down in the perpendicular direction. We also use a power-law cooling process
throughout the flow. We see that a Keplerian disk is produced 
on the equatorial plane, but it is truncated at a place where the sub-Keplerian component 
forms the shock-wave. A CENBOL is produced and the outflows are mostly produced from this region.

We divide the paper as follows: In the next section, we present the model equations for the viscous
flows with cooling. In Section 3, we briefly discuss the methodology of numerical simulations. 
In Section 4, we present the results. Finally, in Section 5, we draw conclusions.

\section{Model Equations including Viscous flows and Power-law Cooling}

The basic equations describing a two-dimensional axisymmetric flow around a Schwarzschild 
black hole are already described in Molteni, Ryu \& Chakrabarti (1996; hereafter MRC96).   
The self-gravity of the accreting matter is ignored. Cylindrical coordinate
$(r, \phi, z)$ is adopted with the z-axis being the rotation axis of the disk. 
We use the mass of the black hole $M_{BH}$,
the velocity of light $c$ and the Schwarzschild radius $r_g=2GM/c^2$ as the
units of the mass, velocity and distance respectively. The equations governing 
the viscous flow have been presented in Paper II in great detail and we do not repeat them here. 

As far as the cooling is concerned, ideally one needs to use Comptonization. However, the process 
is highly non-linear and non-local. On the other hand, the bremsstrahlung cooling which could 
be computed from the local density and temperature is too weak to have any significant effect. 
Therefore, we choose a power law cooling in the energy equation having a temperature dependence 
as $T^\beta$. Hence the cooling rate is ${{\Lambda}_{powcool}} \propto {{\rho}^2}T^{\beta}$, 
where, $\beta > 0 $ is the cooling index. The energy equation becomes:
$$
{{\partial {(\rho {\epsilon})}} \over \partial t} + {{\nabla}.{(\rho {\epsilon} {\bf v})}}
 + {\Lambda}_{powcool} = 0,          \eqno(1)
$$
where, $ {\epsilon} = {p\gamma \over (\gamma-1)}+{{(v_x^2+v_{\theta}^2+v_z^2)}/2} 
+ g, $ is the specific energy, $\gamma$ is the adiabatic index, $\rho$ is the mass density.
Here, ${\Lambda}_{cool}$ is the expression for power-law cooling. So, the energy conservation 
equation of Paper II becomes,
$$
{\partial E \over \partial t} + {1 \over r}{\partial {(E+p)rv_r} \over \partial r}
+ {\partial {(E+p)v_z} \over \partial z} =
- {\rho \left(rv_r+zv_z\right) \over  2\left(\sqrt{r^2+z^2}-1\right)^2\sqrt{r^2+z^2}}
 -{{\Lambda}_{powcool}}. \eqno(2)
$$
Here, energy density $E$ (without the potential energy) is defined as,
$E=p/(\gamma-1)+\rho(v_r^2+v_{\theta}^2+v_z^2)/2$, $\rho$ is the mass density,
$\gamma$ is the adiabatic index, $p$ is the pressure, $v_r$, $v_\theta$ and $v_z$
are the radial, azimuthal and vertical components of velocity respectively.
The default power-law for bremsstrahlung cooling is obtained by taking cooling
index $\beta = {1 \over 2}$.
In an electron-proton plasma, the expression for bremsstralung cooling process (Lang, 1980) is given as,
$$
{\Lambda}_{brems} = 1.43 \times {10^{-27}} N_e N_i {T^{1 \over 2}} {Z^2} {g_f}   
{\bf \it {erg}  {cm^{-3}}  {s^{-1}}} ,  \eqno(3)
$$
where,
$$
{N_i}Z = {\rho \over {(m_p + m_e)}} \approx {\rho \over {m_p}}, \eqno(4) 
$$
i.e.,
$$
{\Lambda}_{brems} = 1.43 \times {10^{-27}}{{\rho}^2}{T^{1 \over 2}} {g_f}, \eqno(5)
$$
where, $m_p$ is the mass of proton, $T$ is the temperature, $g_f$ is the Gaunt
factor. In our work, to increase the cooling efficiency, 
we have taken the cooling index $\beta = 1$. Other constants are as in Eq. (5). So the cooling term in
Eq. 2 reduces to,  
$$
{\Lambda}_{powcool} = 1.43 \times {10^{-27}}{{\rho}^2}{T} {g_f},        \eqno(6)
$$
where, everything is expressed in the CGS units and $g_f$ is the Gaunt factor which is assumed 
to be $1.0$ in our work. The temperature $T$ can be easily obtained from the density ($\rho$) and
pressure ($p$) which are being calculated in our simulation. Using the ideal gas equation, we 
get for the temperature $T$,
$$
T = {p \over \rho}{{\mu {m_p}} \over {k_b}},  \eqno(7)
$$
where, $k_b$ is the Boltzmann constant. We assume, $\mu = 0.5$, for purely hydrogen gas. 

In Paper II, we  already presented the results for a constant viscosity parameter $\alpha$  
in the entire flow as is in vogue in the subject. In the present paper, 
our viscosity parameter is high on the equatorial plane and low, away from from it. 
This is because, as is well known in the case of the models of the dwarf novae outbursts, 
the high viscosity on the equatorial plane actually drives the accretion (e.g., Cannizzo et al. 1982, 1995).
Thus the rate of transport of angular momentum on the 
equatorial plane should be the highest. Away from the plane, the pressure falls very slowly, and thus in 
order to reduce viscous effects, $\alpha$ itself must go down. In the present work, 
instead of using the same $\alpha$ for the whole $x-z$ plane, we choose a smooth the distribution as,
$$
\alpha = {\alpha}_{max} - [{\alpha}_{max}{({z \over {r_{max}}})^{\delta}}],  \eqno(8)
$$
where, $r_{max} = 200, 0 \leq z \leq 200$ and $\delta > 0$. In our cases, we have chosen 
$\delta = 1.5$. Clearly, when $z = 0$, i.e. at equatorial plane, $\alpha = {\alpha}_{max}$, 
while, $\alpha = 0$ for $z = z_{max} = r_{max}$. In order that a Keplerian disk may form on the 
equatorial plane, ${\alpha}_{max}$ must be larger than critical viscosity $\alpha_{crit}$. 
If turbulence is the major source of viscosity, then, clearly 
it will be highest on the equatorial plane. But a precise knowledge is required to get the distribution
of how alpha actually falls with height. That is why we assume a generic distribution.
Since we have no preferred choice one way or the other,
we ran our code with several distributions of the similar nature, but the basic result
was found to remain the same.  

\section{Methodology of the Numerical Simulations}

The setup of our simulation has been described in Paper I and Paper II. However, we briefly
describe it again for completeness. We consider an axisymmetric flow moving in a Pseudo-Newtonian gravitational 
field of a point mass $M_{bh}$ located at the centre in cylindrical coordinates $[r,\theta,z]$. 
The gravitational field can be described by Paczy\'{n}ski \& Wiita (1980) potential,
$$
\phi(r,z) = -{GM_{bh}\over2(R-r_g)}, \eqno{(9)}
$$
where, $R=\sqrt{r^2+z^2}$.

The computational box occupies one quadrant of the x-z plane with $0 \leq r \leq 200$ 
and $0 \leq z \leq 200$. We use the reflection boundary condition on the equatorial plane
and z-axis to obtain the solution in other quadrants. The incoming gas enters the box through the outer boundary
(which is of a vertical cylinder in shape), located at $x_b = 200$. We supply the radial velocity $v_r$,
the sound speed $a$ (i.e., temperature) of the flow at the boundary points. 
We take the boundary values of density from the standard vertical equilibrium 
solution (C89). We scale the density so that the incoming gas 
is ${\rho}_{in} = 1.0$. In order to mimic the horizon of the black hole at one
Schwarzschild radius, we place an absorbing inner boundary at $R = 2.5 r_g$, 
inside which all material is completely absorbed into the black hole. The inner sonic 
point is formed around this radius anyway, so this choice of inner boundary does not
affect the flow dynamics. We fill the grid with the background matter of very low density ${\rho}_{bg} = 10^{-6}$ 
having a sound speed (or, temperature) to be the same as that of the incoming
matter. Hence, the incoming matter has a pressure  $10^6$
times larger than that of the background matter. This matter is put to avoid
unphysical singularities caused by `division by zero'. Of course,  this initial condition is totally  washed out
and replaced by the incoming matter in a dynamical time scale. 
Initially, the values of radial ($v_r$), rotational ($v_\phi$)  
and azimuthal ($v_z$) components are all chosen as zero for all 
the grids except those on the outer boundary. Thus, the Mach number 
is zero everywhere at the beginning of the simulation.
The calculations were performed with $512 \times 512$ cells, so
each grid has a size of $0.39$ in units of the Schwarzschild radius. 
In this work, we are interested to show (a) the formation of CENBOL (at tens of Schwarzschild radii) and 
(b) the formation of a Keplerian disk (few Schwarzschild radii thick). 
So, the resolution which we have ($ \sim 0.4$ Schwarzschild radii)
is enough to catch these salient features. For detailed study of turbulent cells, we require to refine the 
grids. This will be done in future.
All the simulations have been carried out assuming a stellar mass black hole $M_{BH}= 10{M_\odot}$. 
The conversion of our time unit to physical unit is $2GM_{BH}/c^3$, and thus 
the physical time for which the programme was run would scale with the mass of the black hole.
We typically find that the  infall time from the outer to the inner boundary is about $\sim 0.5$s. This is  
computed from the sum of $dr/<v_r>$ over the entire radial grid, $<v_r>$ being averaged over $20$ vertical grids. 
We carry out the simulations for several hundreds of dynamical time-scales. 

\section{Simulation Results}

The total variation diminishing (TVD) method which we use here was initially developed
to deal with laboratory fluid dynamics (Harten, 1983). In the astrophysical context,
Ryu et al. (1994 \& 1996) developed the TVD scheme to study non-viscous astrophysical flows without cooling.
In Paper I, the oscillation phenomenon in non-viscous accretion flows around black holes
related to the QPOs were reported. In Paper II, we studied viscous accretion flows around
black holes using the turbulent viscosity prescription which is given by Shakura \& Sunyaev (1973).
Here, it was shown that the centrifugal pressure supported shocks became weaker with the inclusion 
of viscosity. For sufficiently large viscosity, the shocks disappear altogether 
and the flow becomes subsonic and Keplerian everywhere (except in a region close to the 
horizon). In the present paper, we implemented the distributed 
$\alpha$ (Eq. 8) and the power-law cooling in the code.

We assume the flow to be in vertical equilibrium (C89) at the outer boundary.
The injection rate of the momentum density is kept uniform throughout the injected height 
at the outer edge. We inject through all the radial grids. We stop the simulations 
at $t=95$s (physical time). This time is more than two hundred times the dynamical time of the flow. 
Thus, the presented solutions are at a time long after the transient phase which is over in about a second. 
The results of the simulation are discussed below.
\begin{figure}
\begin{center}
\includegraphics[height=14truecm,width=11truecm,angle=0]{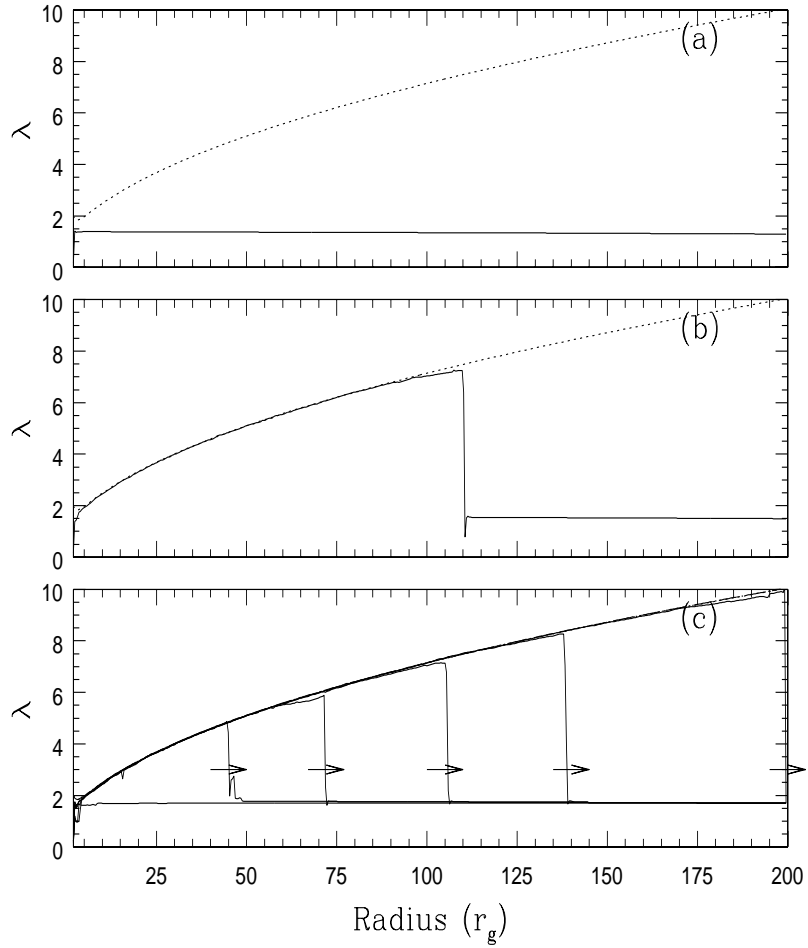}
\caption{Comparison of the specific angular momentum distributions (solid curves) with injected
(a) $\lambda=1.3$, (b) $1.5$ and (c) $1.7$ when the viscosity and the cooling effects are 
included. The results are compared with the Keplerian angular momentum distribution (dotted curves).
Note that, in (b), the Keplerian disk reaches till $\sim 110r_g$. In (c), however, the Keplerian disk
reaching towards the grid boundary. The vertical boundaries are at $t=23.71$s, $35.57$s, 
$47.43$s, $73.34$s and $95$s respectively.} 
\end{center}
\end{figure}

First, we perform two-dimensional numerical simulations with viscosity and cooling
for different values of specific angular momentum $\lambda=1.3$, $1.5$ and $1.7$. 
In the absence of viscosity, the angular momentum would have remained the same as those of the 
injected values. In Fig. 1(a-c), we show the distributions of specific angular momenta (solid 
curve) with Keplerian angular momentum distribution (dotted curve) on the equatorial plane when 
viscosity and cooling are added. The specific energy ($\cal E$) of the 
flow at the equatorial plane ($z = 0$) was chosen to be $0.001$ at the outer boundary.
The viscosity parameter was chosen to be, ${\alpha}_{max} = 0.012$ and the cooling index was 
chosen to be $\beta = 1$. In all the cases, the transient behaviours are over within  $1$s. In (a), where, 
$\lambda = 1.3$, the angular momentum transport rate is negligible even when
viscosity is added. But, the transport rate is significant for the intermediate value:
(b) $\lambda=1.5$. For larger (c) $\lambda=1.7$, the angular momentum has been transported rapidly and
the distribution of specific angular momentum in (c) coincides with the Keplerian value.
In Fig. 1c, from left to right, we have plotted shock transitions
at $ t = 23.7$s, $t = 35.6$s, $t = 47.4$s, $73.3$s and $95$s respectively. 
In (a) , the shock does not exist for such a low angular momentum flow (C89). 
So the case will never produce any Keplerian disk also.
Both (b) and (c) both produce shocks as they have sufficient angular momentum.
However, while in (b), the shock remains standing 
for ever, in Fig. (c), the shock propagated outward and the whole disk becomes Keplerian. 
The reason is that $\alpha < \alpha_{crit}$ in (b), and $\alpha < \alpha_{crit}$ in (c).
In (c), if we run longer, the shock will propagate to large distance making
the whole post-shock region a Keplerian disk. Only region between the horizon and the inner sonic point ($ \sim 2.5 r_g$)
will remain supersonic and sub-Keplerian.

\begin{figure}
\begin{center}
\includegraphics[height=12truecm,width=8truecm,angle=0]{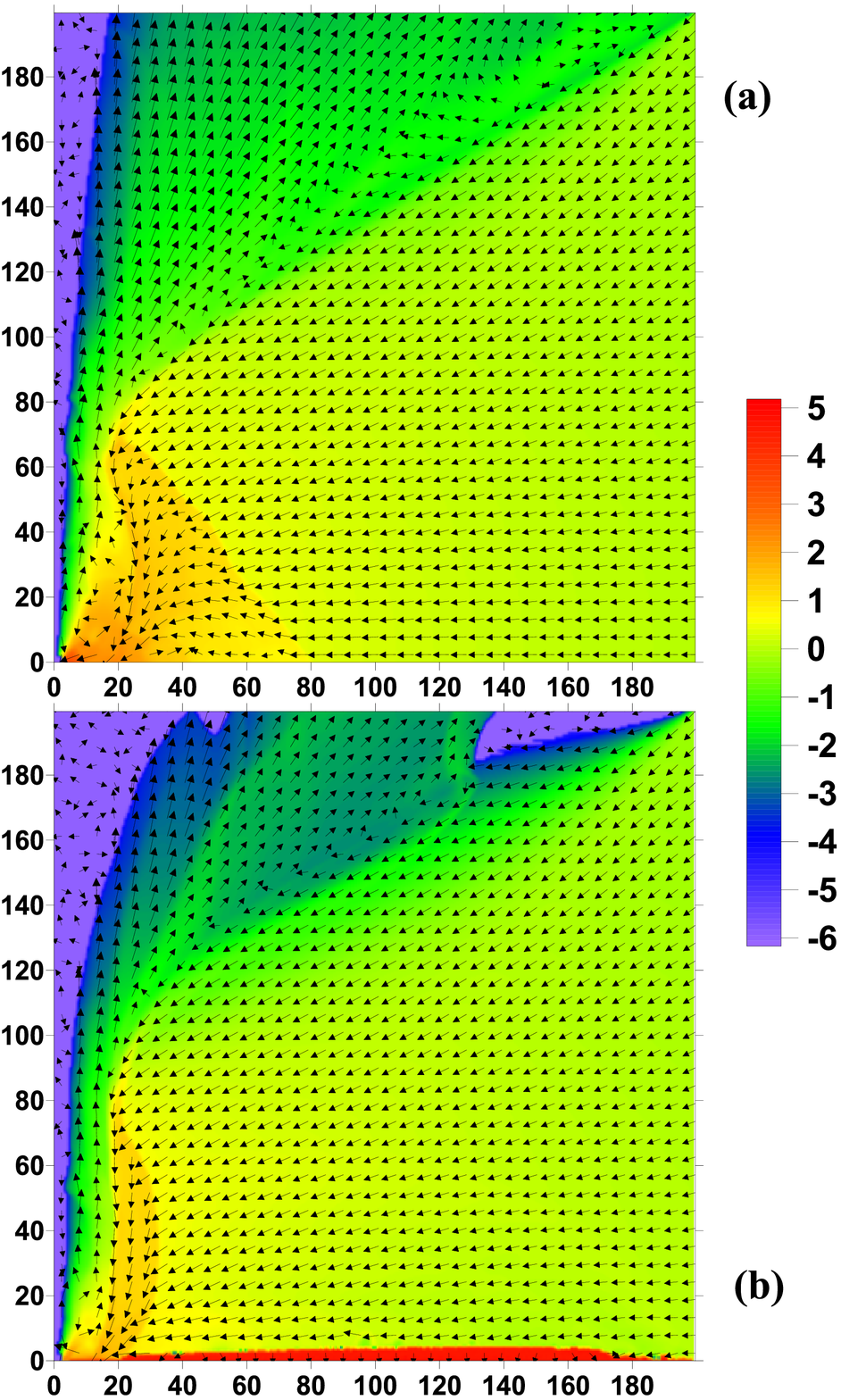}
\caption{Changes in the density and velocity distributions  at $t= 95$s (a) without and (b) with
the inclusion of viscosity and cooling.
Densities in normalized units are plotted in logarithmic scale as in the scale on the right.
The density ranges from ${{log}_{10}}\rho = -6$ to $5$ in both the Figures.
A two component flow is clearly formed in (b).}
\end{center}
\end{figure}

\begin{figure}
\begin{center}
\includegraphics[height=12truecm,width=8truecm,angle=0]{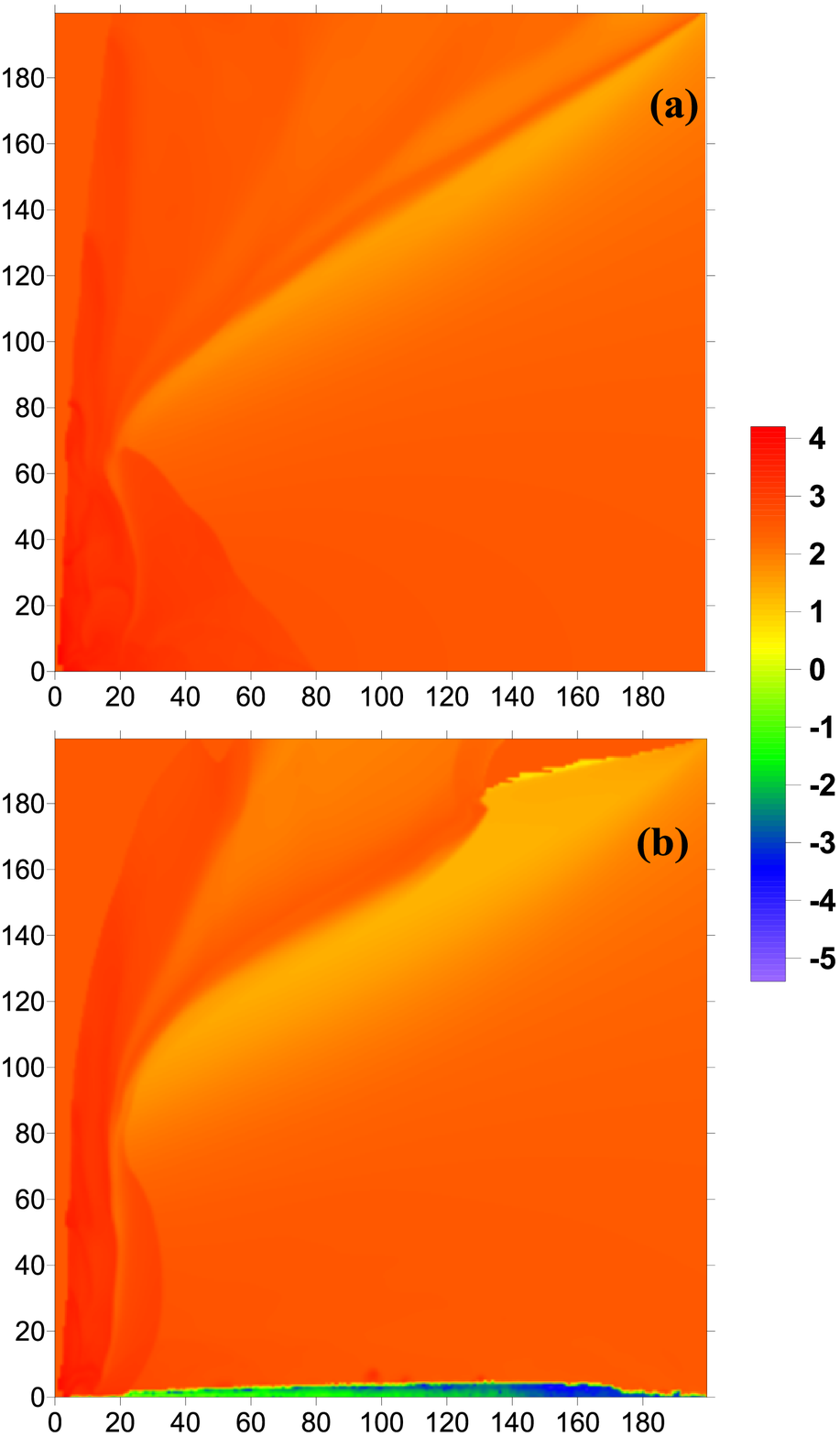}
\caption{Temperature distribution with logarithmic scale for both non viscous and 
viscous cases. For both cases, the temperature ranges from ${{log}_{10}}T = -5$  to $4.1$.} 
\end{center}
\end{figure}
Fig. 2(a-b) shows the velocity and the density distribution of the flow (a) without viscosity and cooling 
and (b) with viscosity and cooling. In order to have a meaningful comparison, 
all the runs were carried out up to $t=95$s. For both the cases, $\lambda = 1.7$
and ${\cal E} = 0.001$ were chosen. In Fig. 2b, we take ${\alpha}_{max} = 0.012$ and $\beta = 1$. 
The density distributions in Fig. 2 are plotted in the logarithmic scale shown on the right. 
We note that on the equatorial plane, a Keplerian disk has formed out of sub-Keplerian matter
by transporting angular momentum outward.
Close to the outer boundary, near the equatorial plane, we are injecting sub-Keplerian matter 
and thus the Keplerian disk is disrupted there. In a realistic situation, the Keplerian 
would continue till the outer boundary. In Fig. 3(a-b), we show the temperature distributions 
in keV as per color (logarithmic) scale on the right. In the absence 
of cooling and viscosity, in Fig. 3a, the single component sub-Keplerian flow forms. In Fig. 3b,
because of higher viscosity, the flow has a Keplerian distribution near the equatorial region. 
Because of the cooling effect, the region with a Keplerian distribution
is cooler and denser. Comparatively low density sub-Keplerian matter stays away from the equatorial plane. 
For both the cases, the Centrifugal Pressure supported BOundary Layer (CENBOL) 
forms. Since the inner boundary condition on the horizon forces the flow to be sub-Keplerian 
irrespective of their origin (Chakrabarti 1990, 1996), the Keplerian and sub-Keplerian matter 
mixes (in CENBOL) before entering a into a black hole as a single component sub-Keplerian flow. 
A very thin barely resolved Keplerian component continues on the equatorial plane till the inner sonic point.

It is pertinent to compare our results with the results of some of the previous studies.
Igumenshchev \& Abramowicz (1999, 2000), Stone, Pringle \& Begelman (1999),
Proga \& Begelman (2003) do not find the existence of shocks.
In those simulations, the authors mainly concentrated on the time evolution injected 
Keplerian or almost Keplerian disk. As the Keplerian flow itself is a subsonic flow, 
the time evolution of the Keplerian flow  will not produce a shock. 
Those simulations were also viscous, so a Keplerian flow remained Keplerian until the inner sonic point.
On the contrary, our flows are advective with a significant radial velocity component 
and the flow has an angular momentum much less than that of the Keplerian disk. With addition of 
viscosity the distribution becomes Keplerian along the equatorial plane
and the cooling  ensures that this Keplerian flow indeed radiates like a standard Keplerian disk. 
Since above and below the equatorial plane, the density is very low, the viscosity and cooling
are inefficient, the flow remains sub-Keplerian and produces shocks. So
we have a complete solution in which the equatorial part behaves like a Keplerian disk, while the other
part away from the equatorial plane behaves like a transonic flow with shock waves.

It is evident that the out flows are generated from inflow at both Fig 2a and 2b. 
The centrifugal barrier (CENBOL) produces shock, and the post-shock flow become hot.
Due to the heating at the jet-base (CENBOL) 
and the subsequent expansion in the vertical direction, the outflow is generated from the disk.
The excess thermal gradient force along the vertical direction in the post-shock flow drives a part 
of the accreting matter as the bipolar outflow which is believed to be the predecessor of the observed jet.
We thus have the indication that the shock heating is an important ingredient in the ejection of matter from the 
disk surface. Indeed in presence of cooling, the outflow is quenched (Garain et al. 2012). 
The formation of such out flows has been shown previously in hydrodynamical simulations (MLC94, MRC96, Paper I, Paper II).
Most recently, similar types of outflows from inflow is also observed in Yuan et al. (2012).
\begin{figure}
\begin{center}
\includegraphics[height=12truecm,width=12truecm,angle=0]{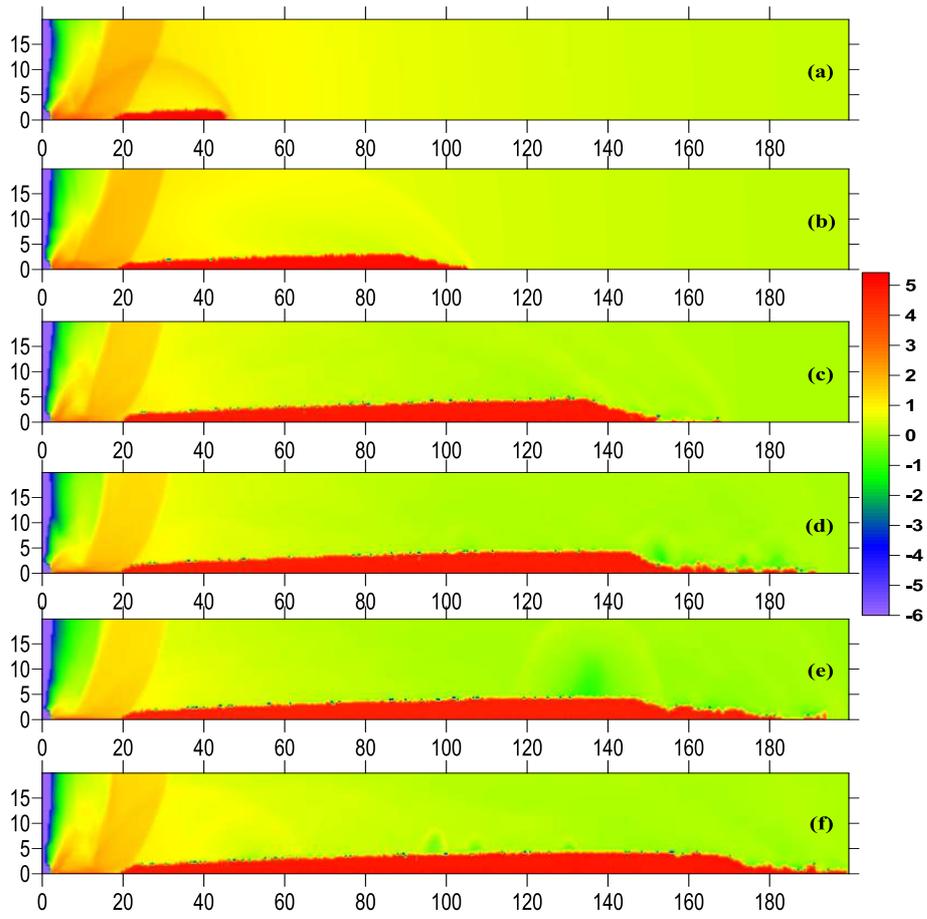}
\caption{Time variation of density distribution (in logarithmic scale) at six different times: 
$t = 20, 50, 70, 80, 90$ and $95$s. The density ranges from ${{log}_{10}}\rho = -6$ to $5.2$. 
The CENBOL forms at $r \sim 20$.} 
\end{center}
\end{figure}

A standard Keplerian disk forms in two steps: (a) The viscosity must be sufficiently
high ($\alpha > {{\alpha}_{crit}}$ as discussed in Paper II) to produce a Keplerian distribution and
(b) The cooling must be sufficiently high to emit a black body locally. 
In order to show how the Keplerian disk actually forms out of the sub-Keplerian flows, 
we zoom the region close to the equatorial plane and show the results at six different 
times: $t = 20, 50, 70, 80, 90$ and $95$s. Different colours correspond to different 
densities as marked in the scale on the right. The red (online version) colour
corresponds to $\sim 10^5$ while the yellowish green corresponds to $\sim 1$, the injected density. 
The CENBOL is at $r \sim 20$ where the density increases the factor of a few. Injected matter near 
the equatorial plane being strongly sub-Keplerian, it disrupts the Keplerian disk in the region $r\sim 180-200$. 
In an astrophysical context such disruption will not take place.
Note that the outer front of the Keplerian disk moves very slowly as compared to the inflow velocity of the sub-Keplerian
matter. In the frame of the sub-Keplerian matter the Keplerian disk behaves as an obstacle. This causes 
the formation of an wake at the tip of the outer edge of the Keplerian flow.
If one starts with a Keplerian disk and gives no viscosity, the disk will immediately 
accrete and it will not remain Keplerian.  So a disk can be Keplerian even without cooling. 
But a disk cannot be remain Keplerian without sufficient viscosity.

\begin{figure}
\begin{center}
\includegraphics[height=10truecm,width=10truecm,angle=0]{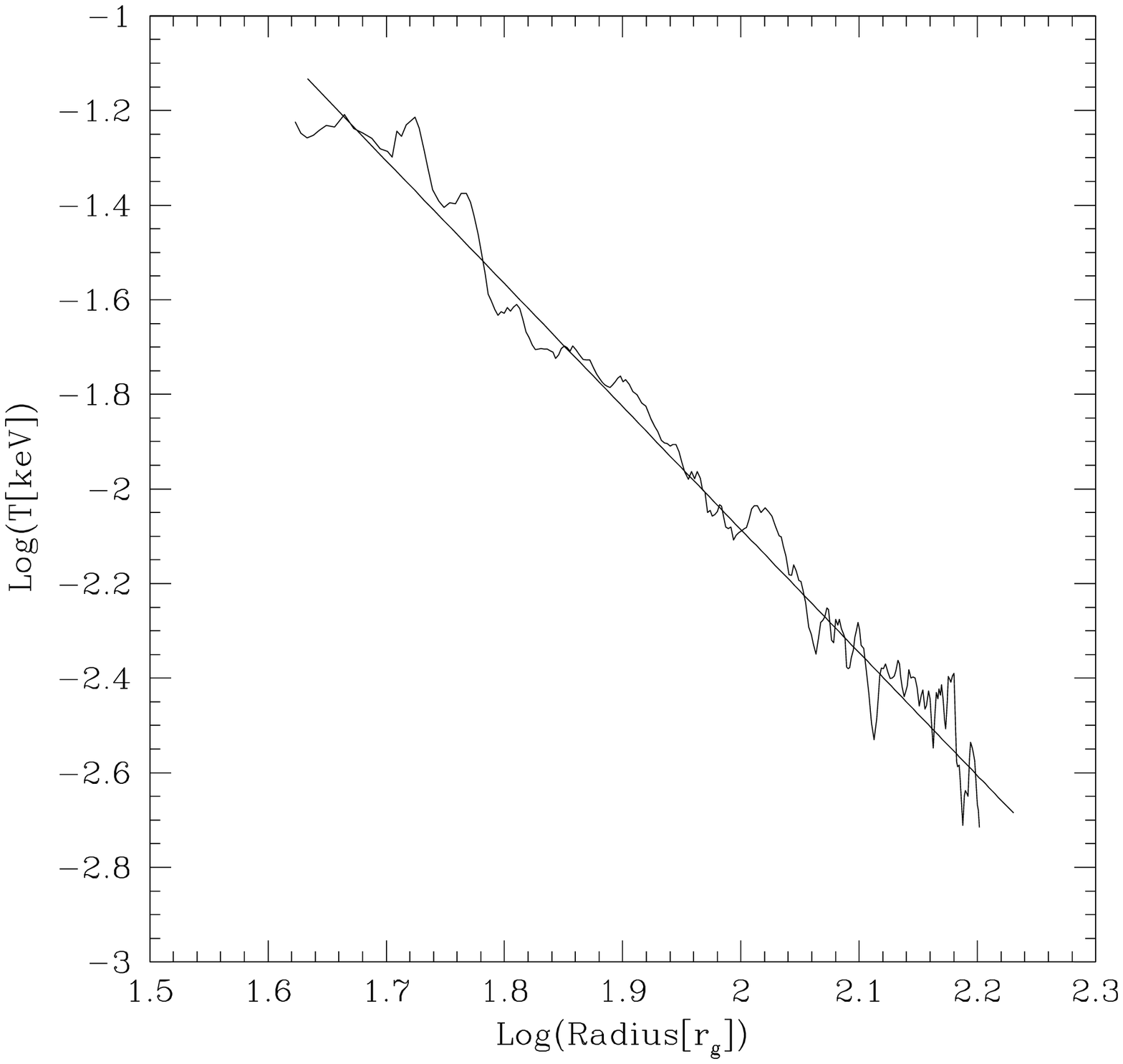}
\caption{Log-Log plot of the radial distribution of time averaged Temperatures (in KeV) of the Keplerian component 
at $t=95$s. The slope $\gamma$ of the power law distribution $T(r) \sim r^{-\gamma}$ is $2.6$.} 
\end{center}
\end{figure}

We have already shown in Fig. 1 that the angular momentum distribution of the flow close to the equatorial plane is
indeed Keplerian as far as the Keplerian component goes. Now we wish to show how the temperature is distributed. In the 
standard Keplerian disk (Shakura \& Sunyaev, 1973), in the optically thick regime, the black body cooling
law $\propto T(r)^4$ leads to the disk temperature distribution of $T(r) \sim r^{-3/4}$. However,
the cooling we employed here is $\propto T(r)^\beta$, where $\beta=1$. This leads to a possible
distribution of $T(r) \sim r^{-3}$, steeper than a Keplerian disk emitting black body. In Fig. 5, we plot 
the radial distribution of vertically averaged temperature at $t = 95$s. We have plotted between 
$30 r_g$ and $150 r_g$ to avoid the boundary effects. The temperature just outside CENBOL at $r\sim 20 r_g$ 
is found to be around $0.1$ keV. The slope of the distribution ($T(r) \sim r^{-\gamma}$) is found to be $\gamma = 2.6$. 
This converged result is close to our predicted value. In future, when we add the Comptonization scheme inside the simulation, 
we anticipate that $T(r) \sim r^{-3/4}$ distribution would be achieved.

We now carry out a numerical experiment to show the `hysteresis' effect between 
the formation and the destruction of the Keplerian disk. In Fig. 6(a-d), we show results 
of the simulations (with parameters same as in Fig. 4) at (a) $t=23.7$s  and (b) $t=59$s 
respectively. The density scale is shown on the right (logarithmic). With the output of (b).
we remove the viscosity and cooling and ran again for another $59$s. In Fig. 6c, we show the results at
$t=95$s and in Fig. 6d we show the results at $t=118$s. We note that turbulent cells are produced
all over the flow and the flow pattern is greatly disturbed. Most interestingly, if there were no hysterisis
effects, the Fig. 6c and Fig. 6a would have been alike and Fig. 6d and Fig. 2a would have been alike. 
However, they are not. Naturally, the spectrum would also have the same effect. Clearly, we need to carry 
the simulation for a much longer time to bring back the CENBOL in the inviscid flow as in Fig. 2a. 
In the outburst sources (Mandal \& Chakrabarti, 2010) exactly the same hysterisis effect could be seen. 
The time variation of the Keplerian and the sub-Keplerian flows do not obey the same route in the 
onset and the decline phases of the outbursts. This will be elaborated elsewhere.

\begin{figure}
\begin{center}
\includegraphics[height=13truecm,width=13truecm,angle=0]{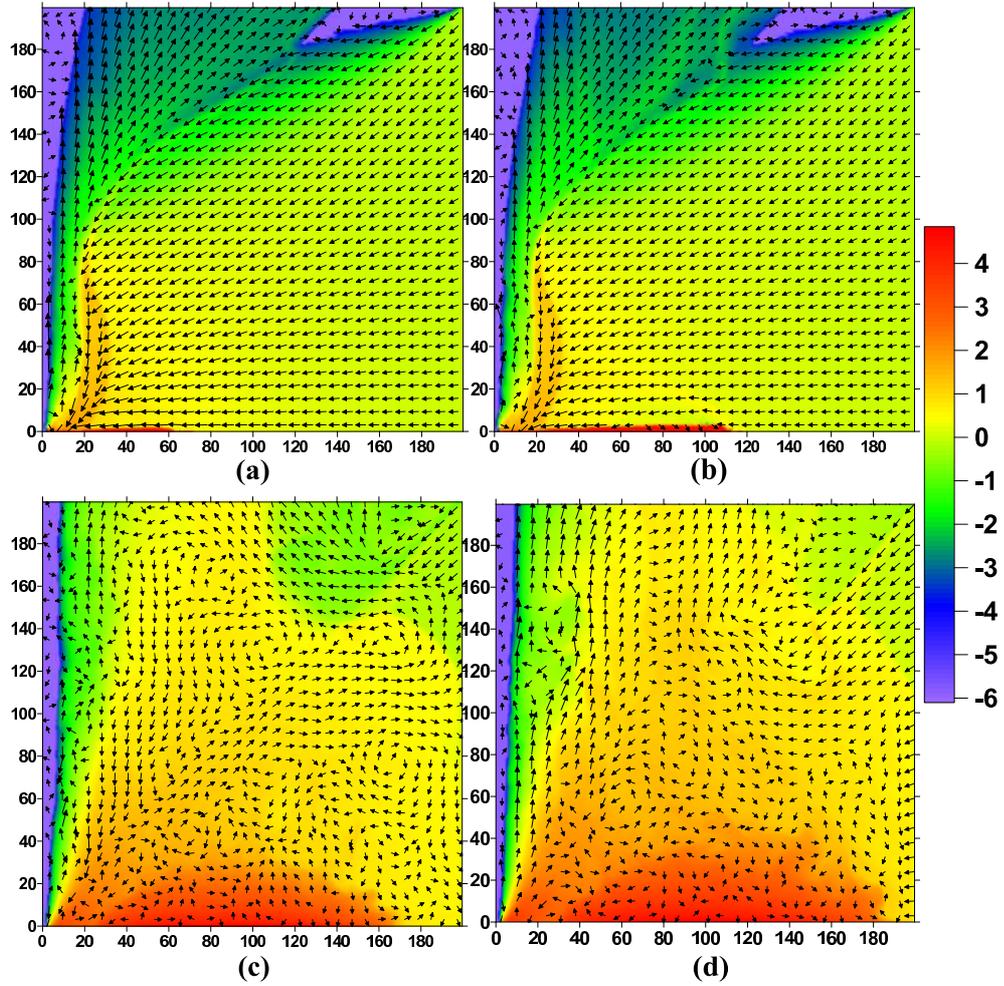}
\caption{(a-b) Formation of a Keplerian disk from the injected sub-Keplerian flow in
presence of viscosity and cooling effects: (a) at $t=23.7$s and (b) at $t=59$s. (c-d) Evaporation 
and mixing of the Keplerian disk started after removal of cooling and viscosity : (c) at $t=95$s and (d)$t=118$s. 
For all cases, density ranges from ${{log}_{10}}{\rho} = -6$  to $5$.}
\end{center}
\end{figure}

\section{Discussions}

Two Component Advective Flow (TCAF) solution in the black hole astrophysics was formulated from the theoretical study 
of the behaviour of topology of viscous flows around black holes. One component having a higher viscosity
is a cooler, Keplerian disk on the equatorial plane. The second component is a low angular momentum and
low viscosity flow which forms a standing or oscillating shock. The region between the shock and the 
inner sonic point behaves as a boundary layer (the Centrifugal pressure supported boundary layer or CENBOL). 
Here the kinetic energy of pre-shock matter is converted to thermal energy. The flow is puffed 
up and forms a geometrically thick disk which behaves as the hot Compton cloud. The CENBOL is the region which 
produces and initially collimates the outflows or jets. The spectrum is softer when the CENBOL is smaller or non-existant.  
The spectrum is harder when the CENBOL is present. When the CENBOL oscillates, low-frequency QPOs form. Furthermore,
it is the only configuration which arises out purely from theoretical considerations. So it is important
to prove that not only the TCAF configuration is realizable, it is also a stable system.

So far, there was no numerical simulations in the literature to show that TCAF solution is realizable as a whole. 
and there was no simulation to show whether such a configuration is at all stable.
It is to be noted, that in the literature, studies have been made on Bondi flows and various accretion disks
(e.g., Hawley et al. 1984a,b; Molteni, Sponholz \& Chakrabarti, 1996;  Igumenshchev \& Abramowicz, 1998, 2000), not on 
the formation of TCAF which is basically a combination of the
generalized Bondi flow with a Keplerian disk in the pre-shock region, 
and a thick accretion disk with an outflow in the post-shock region. Our result, for 
the first time, shows that if one assumes that the viscosity is maximum on the equatorial plane, then, 
a low-angular momentum injected flow is converted into a TCAF. We show that the injected 
flow segregated in the Keplerian and the sub-Keplerian components.  
The sub-Keplerian component produced a shock at around $20r_g$ (This depends on the angular 
momentum of the injected flow) and the resulting CENBOL did not allow the Keplerian 
disk to have the normal structure up to innermost stable circular orbit or ISCO.
The Keplerian flow remained extremely thin (about a grid thick) inside the CENBOL. 
However, for $r\geq 20r_g$, the density of the flow becomes very high and the temperature 
becomes very low. The slope of the temperature distribution is in line with our choice of $\beta$.
We have also shown that there is clearly a hysterisis effect in that, the time taken to 
form a Keplerian disk upon introduction of heating and cooling is faster that the time it takes to  
return back to the original inviscid configuration. This is because the cooler matter of the 
Keplerian flow has lesser thermal drive to fall in. It is possible that the hysterisis 
effects seen in the outburst sources in the onset (turning on the viscosity) and the decline 
(turning off the viscosity) phases is precisely due to this effect: the formation and the disappearance of the 
Keplerian flow takes different times.

So far, we have captured all the salient features of the Keplerian disk, by introducing a power-law cooling 
effect. In order to produce an exact standard disk which emits multi-colour black body as well,  
we need to include the radiative transfer problem {\it ab initio}. We need to generate photons using bremsstrahlung
and scatter them by the Keplerian (angular momentum) component to obtain black body radiation
self-consistently. The emitted photons would then scatter from the CENBOL and the outflows and 
produce harder radiations, observed in black hole candidates. In future, we plan to carry out this analysis. 

{}
\end{document}